\documentclass[aps,pra,twocolumn,noshowpacs,superscriptaddress,preprintnumbers, kamsmath,amssymb]{revtex4}

\usepackage{graphicx, epsfig}
\usepackage{amssymb, amsmath, amsfonts, amsthm}
\usepackage{multirow}
\usepackage{dcolumn}

\usepackage[utf8]{inputenc}

\DeclareMathAlphabet{\mathbbm}{U}{bbm}{m}{n}
\SetMathAlphabet\mathbbm{bold}{U}{bbm}{bx}{n}
\DeclareMathAlphabet{\mathbbmss}{U}{bbmss}{m}{n}
\SetMathAlphabet\mathbbmss{bold}{U}{bbmss}{bx}{n}
\DeclareMathAlphabet{\mathbbmtt}{U}{bbmtt}{m}{n}

\newcommand{\bra}[1]{\ensuremath{\left\langle #1 \right\vert}}
\newcommand{\ket}[1]{\ensuremath{\left\vert #1 \right\rangle}}
\newcommand{\op}[1]{\ensuremath{\boldsymbol{\mathsf{\hat{#1}}}}}

\newcommand{\braket}[2]{\ensuremath{\left\langle #1 | #2 \right\rangle}}
\newcommand{\ketbra}[2]{\ensuremath{\left|#1\right\rangle\kern-3pt\left\langle #2 \right|}}
\newcommand{\opBraket}[3]{\ensuremath{\left\langle #1 \left| #2 \right| #3 \right\rangle}}
\newcommand{\daggered}{\ensuremath{^\dagger}}

\newcommand{\absSq}[1]{\ensuremath{\left|#1\right|^2}}
\newcommand{\dd}{\ensuremath{\, \textnormal{d}}}

\newcommand{\unity}{\ensuremath{\mathbbm{1}}}

\newcommand{\trap}{\operatorname{trap}}

\newcommand{\ia}{\operatorname{int}}
\newcommand{\diag}{\operatorname{diag}}

\newcommand{\addunit}[2]{\newcommand{#1}{\ensuremath{\text{#2}}}}
\newcommand{\reciprocal}[1]{\ensuremath{\text{#1}^{\text{-1}}}}

\addunit{\kilo}{k}
\addunit{\nano}{n}
\addunit{\mega}{M}
\addunit{\volt}{V}
\addunit{\meter}{m}
\addunit{\au}{a.u.}
\addunit{\centi}{c}
\addunit{\pico}{p}
\addunit{\hertz}{Hz}
\addunit{\second}{s}
\newcommand*{\wavenumbers}{\centi\reciprocal\meter}

\begin{document}


\title{The Quantum Speed Limit of Optimal Controlled Phasegates \\
       for Trapped Neutral Atoms}

\author{Michael H. Goerz}\email{goerz@physik.uni-kassel.de}
\affiliation{Institut für Theoretische Physik, Freie Universität Berlin,
  Arnimallee 14, D-14195 Berlin, Germany}
\affiliation{Institut f\"ur Physik, Universit\"at Kassel,
  Heinrich-Plett-Str. 40, D-34132 Kassel, Germany}
\author{Tommaso Calarco}
\affiliation{Institut für Quanteninformationsverarbeitung, Universität Ulm,
  D-89069 Ulm, Germany}
\author{Christiane P. Koch}\email{christiane.koch@uni-kassel.de}
\affiliation{Institut für Theoretische Physik, Freie Universität Berlin,
  Arnimallee 14, D-14195 Berlin, Germany}
\affiliation{Institut f\"ur Physik, Universit\"at Kassel,
  Heinrich-Plett-Str. 40, D-34132 Kassel, Germany}

\date{\today}

\begin{abstract} 
  We study controlled phasegates for ultracold atoms in an optical
  potential. A shaped laser pulse drives transitions between
  the ground and electronically excited states where the atoms
  are subject to a long-range $1/R^3$ interaction. We fully account
  for this interaction and use optimal
  control theory to calculate the pulse shapes. This allows us to determine
  the minimum pulse duration, respectively, gate time $T$ that is
  required to obtain high fidelity. 
  We accurately analyze the speed limiting factors, and we find the gate time to be limited either by the interaction
  strength in the excited state or by the ground state vibrational
  motion in the trap. The latter needs to be resolved by the pulses in
  order to   fully restore the motional state of the atoms at the end
  of the   gate. 
\end{abstract}


\maketitle 

\section{Introduction} 

The physical realization of a quantum computer requires the
implementation of a set of universal gates \cite{NielsenChuang}.
The most difficult part is generally the two-qubit gate since
it involves interaction between two otherwise isolated quantum
systems. In proposals for quantum computing with ultracold
neutral-atom collisions \cite{JakschPRL99,CalarcoPRA00},
the two-qubit gate operation involves atomic motional degrees of
freedom \cite{AnderliniNat07,TrotzkySci08},
most often following adiabatic processes. This implies
frequencies much lower than those characteristic of the trap, typically
around a few tens of kHz. When
long-range interactions, like dipole-dipole forces between Rydberg
atoms \cite{JakschPRL00,IsenhowerPRL10,WilkPRL10},
are employed, the relevant energy scales are larger
and gate speed can in principle  reach a few GHz.

Here, we study the limits to the two-qubit gate operation time for
resonant excitation of two ultracold atoms 
into an electronically excited molecular
state. This may be a low-lying state, like those used for
photoassociating two atoms to form a
molecule \cite{KochPRA06a,KochPRA08},  
or high-lying Rydberg states \cite{JakschPRL00,IsenhowerPRL10,WilkPRL10}. 
In this scenario, the system dynamics becomes more complex, involving
motion under the influence of the excited state potential.
A high-fidelity gate can then no longer be designed ``by
hand''. 
Fortunately, since any gate operation  corresponds to a
unitary transformation on the
qubit basis, its implementation can be formulated as a coherent control
problem \cite{PalaoPRL02,PalaoPRA03}.
Solutions to the control problem can be found theoretically within the
framework of optimal control \cite{Tannor92,SomloiCP93,ZhuJCP98}.
Such an approach has been explored theoretically for
molecular quantum computing with qubits
encoded in vibrational states \cite{PalaoPRL02,TeschPRL02}.
Experimentally the control problem is solved using 
femtosecond laser pulse shaping combined with 
feedback loops \cite{JudsonPRL92}.
Implementation of single-qubit gates with shaped picosecond pulses has
recently been demonstrated for a qubit encoded in hyperfine levels of
an atomic ion \cite{CampbellPRL10}. In this experiment, the fundamental
limit to the gate operation time is set by the inverse of the hyperfine
splitting since the hyperfine dynamics are required to realize arbitrary qubit
rotations. The gate duration is in particular much shorter than the
period of atomic motion in the trap \cite{CampbellPRL10}. 

Our goal is to implement a controlled phasegate,
\begin{equation}
  \op{O} = \diag \left( e^{i \chi}, 1, 1, 1 \right),
  \label{eq:target_operator}
\end{equation}
between two qubits carried by neutral atoms. Shaped short laser
pulses are used to drive transitions into an electronic state where
the two atoms are interacting. Obviously, 
the minimum gate operation time will depend on the interaction
strength. A second timescale comes into play because the
interaction couples electronic and nuclear dynamics, inducing 
vibrational excitations of the two atoms. The vibrational period of
the trap might thus also affect the minimum gate operation time.

The limit to how fast a quantum gate can be performed is closely
related to the minimum time it takes for a quantum system to evolve from
an initial state to an orthogonal state, see for example
\cite{LloydNat00} and references therein. This bound has been named the 
quantum speed limit; it is given in terms of the average energy and
energy uncertainty of the quantum evolution \cite{LevitinPRL09}. Since
evaluation of the bound requires knowledge of the full spectrum, one can
typically evaluate it analytically only for simple model
systems. Numerically, the bound can be determined using optimal
control where the breakdown of convergence indicates that
the quantum speed limit has been reached \cite{TommasoPRL09}. 
The following procedure allows us to determine the quantum speed limit
for our desired two-qubit gate: We first set the gate operation time
to a sufficiently large value to obtain a high fidelity implementation
by optimal control. Then we reduce
the gate time until the optimization algorithm cannot anymore find 
a high-fidelity solution. Analysis of the fairly complex system
dynamics reveals which part of the overall system dynamics limits the
gate operation time. 

\begin{figure}[tb]
  \centerline{
    \includegraphics{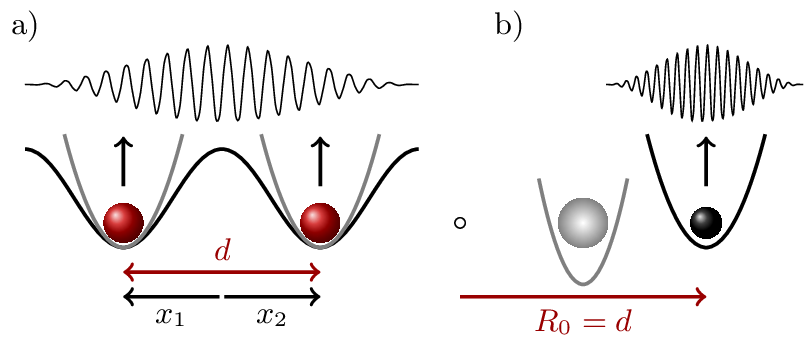}
  }
  \caption{(color online) Two calcium atoms in neighboring sites of an optical
  lattice (a), separated by distance $d$. The lattice potential can be taken in
  a harmonic approximation, allowing to separate the motion of both atoms into
  the center-of-mass motion and the interatomic motion (b). A phasegate is
  implemented by applying a shaped short laser pulse.}
  \label{fig:trapped_atoms}
\end{figure}

We consider neutral atoms trapped in optical tweezers or in
neighboring sites of an optical lattice 
(Fig.~\ref{fig:trapped_atoms}).
The sites to the right and left of the atom pair are assumed to be
empty. We first consider alkaline-earth atoms which possess 
extremely long-lived excited states, cf. Ref. \cite{BoydPRL07} and
references therein. The qubits can therefore be encoded directly in
the electronic states, that is, the atomic ground state $^1S_0$ 
together with the $^3P_1$ clock-transition state form a qubit.
The laser is slightly detuned from the dipole allowed atomic
transition between the $^1S_0$ ground state and the $^1P_0$ excited
state, exciting the atom pair into the $B^1\Sigma_u^+$
molecular state. When both atoms are in the ground state, their
interaction is of van-der-Waals type and practically zero at the trap
distance. Due to different exchange interaction in the electronically
excited state, the $B^1\Sigma_u^+$ state scales as $1/R^3$ at long
range. This interaction may be employed to entangle the qubits
provided the time that the atom pair resides in this state is much
shorter than its lifetime of a few nanoseconds.
The interaction strength is determined by the distance of the
atoms. For realistic lattice parameters, $d \ge 200\,$nm, the
interaction is too weak to generate entanglement in a sufficiently
short time.
On the other hand, we would like to probe what are the factors that limit the
achievable speed on a general level. To this aim, we first explore a regime
that cannot be realized in experiments, but presents a clear separation of
timescales that allows for a clearer interpretation of the dynamics. We start
assuming a fictitious 
distance of $d=5\,$nm and a correspondingly unrealistic atomic trap frequency
of 400 MHz. This allows us
to identify the limiting
factors for fidelity and gate time. Formally, our Hamiltonian is
equivalent to the one yielding a Rydberg phasegate for alkali
atoms~\cite{JakschPRL00}. 
We therefore study in a second step the implementation of a
phasegate based on very strong dipole-dipole interaction in the
excited state for realistic lattice spacings. In particular, we seek
to answer the question whether excitation of the vibrational motion
can be avoided if the excited state interaction is strong enough to
allow for very short gate pulses.

The paper is organized
as follows. Section~\ref{sec:theory} introduces our model for the two
atoms and summarizes how quantum gates can be implemented using optimal
control theory. The numerical results are presented in
Sec.~\ref{sec:results} with Sec.~\ref{subsec:Ca} devoted to generation
of entanglement for two calcium atoms via interaction in the
$B^1\Sigma_u^+$ state. A generic dipole-dipole interaction, $-C_3/R^3$, is
considered in Sec.~\ref{subsec:dipole} where we determine the gate
operation time for varying $C_3$. We draw our conclusions in
Sec.~\ref{sec:concl}.

\section{Theoretical Approach} 
\label{sec:theory}

\subsection{Modeling Two Atoms in an  Optical Lattice}
\label{subsec:model}

\begin{figure}[tb]
  \centerline{
    \includegraphics{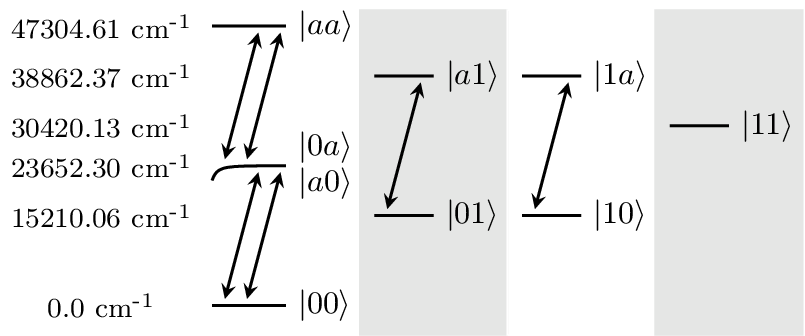}
  }
  \caption{Levels and laser-induced
    transitions for two qubits encoded in electronic
    states of two calcium atoms.}
  \label{fig:2q_levels}
\end{figure}

\begin{figure}[tb]
  \includegraphics{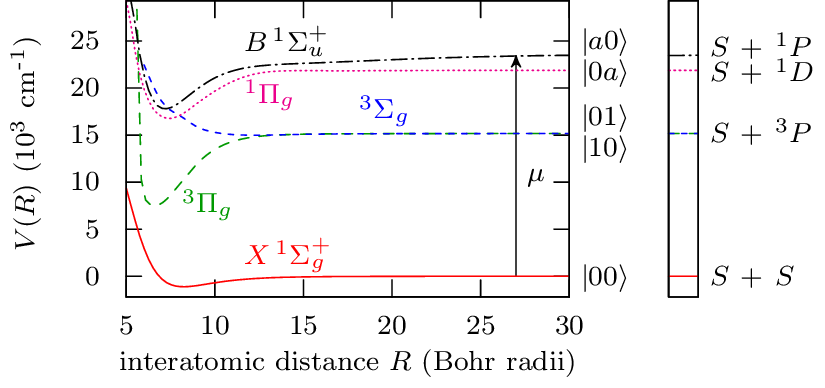}
  \caption{(color online)
    Potential energy curves describing the interaction of two
    calcium atoms in the five lowest
    electronic states. The asymptotic values corresponding
    to $R=\infty$ are indicated on the right.}
  \label{fig:ca2pot}
\end{figure}

We consider the following qubit encoding in a single calcium atom:
The $^1S_0$ ground state corresponds to the qubit state \ket{0} and is
used to define  the zero of 
energy. The $^3P_1$ first excited state, taken to be the qubit state
\ket{1}, then occurs at $E_1 = 15210\,\wavenumbers$. We consider the
$^1P_1$ level as auxiliary state, \ket{a}, with energy $E_a =
23652\,\wavenumbers$. 
For two atoms we then obtain nine electronic states \ket{00}, \ket{01}, \ket{0a},
\ket{10}, \ket{11}, \ket{1a}, \ket{a0}, \ket{a1}, with asymptotic energies
$E_{ij} = E_{i} + E_{j}$ for the states $\ket{ij}$ ($i,j =
0,1,a$). This is depicted in Fig.~\ref{fig:2q_levels}. 
With the laser tuned close to the transition $\ket{0}
\leftrightarrow \ket{a}$, no further electronic states 
will be resonantly populated. 
The potentials describing the interaction between the two atoms in the
five lowest electronic states are shown in Fig.~\ref{fig:ca2pot}. 
The $X^1\Sigma_g^+$ ground state potential corresponding to the
two-qubit state $\ket{00}$ shows a $1/R^6$ behavior at
long range such that the atoms are effectively non-interacting at any
relevant distance, while the auxiliary $B^1\Sigma_u^+$ state corresponding to
$\ket{0a}$ and $\ket{a0}$ goes as $1/R^3$. 
The remaining potentials corresponding to the two-qubit states
$\ket{aa}$, $\ket{a1}$,
$\ket{1a}$, $\ket{11}$,  $\ket{10}$, and $\ket{01}$,  
are also essentially zero at the relevant distances. 
The potentials and transition dipole moment 
functions employed in the following calculations are gathered from Refs.
\cite{BusseryPRA03,BusseryMolPhys06}.
For a Rydberg phasegate with alkali atoms, the qubit encoding is the
standard one in hyperfine levels of the electronic ground state; and
the auxiliary level corresponds to the Rydberg state (we omit in this
description any intermediate level that might be needed for
near-resonant two-photon excitation of the Rydberg state). In this
case, the ground state potential is simply set to zero, while the
auxiliary state is modeled by a generic $C_3/R^3$ potential.

The parameters of the optical lattice are chosen such that the motion
of the atoms is restricted to one spatial dimension ($\omega_\perp
\gg\omega_0$). Approximating the resulting trapping potential for the two atoms
by two displaced harmonic oscillators, we can separate the center of
mass coordinate  which is subsequently integrated out, and
the relative coordinate, i.e. the interatomic distance $R$.

The Hamiltonian of our model, taking into account the nine two-qubit 
states, the motion in the trap, and the interaction between the atoms,
is given by 
\begin{eqnarray}
    \op{H} &=& \op{H}_{1q} \otimes \unity_{1q} \otimes \unity_{R}
    + \unity_{1q} \otimes \op{H}_{1q} \otimes \unity_{R}
    + \op{H}_{\ia} \nonumber \\
    &=& \sum_{i,k} \ketbra{ik}{ik} \left[ \op{T} + \op{V}_{\trap}(R)
      + \op{V}^{ik}_{BO}(R) + E_{ik} \right] \nonumber \\
    &&+ \epsilon(t) \sum_{i \neq j,k}  \big[ \ketbra{ik}{jk} + \ketbra{ki}{kj} \big]
    \otimes \op{\mu}_{ij}(R).
    \label{eq:full_hamil}
  \end{eqnarray}
Here, $\op{H}_{1q}$ denotes the Hamiltonian for
a single three-level atom, 
$\unity_{1q}$ is the identity in $SU(3)$,
and $\unity_{R}$ represents the identity for the motional degree of
freedom. The interaction Hamiltonian $\op{H}_{\ia}$ contains the
Born-Oppenheimer potentials $\op{V}_{BO}$, and the laser pulse
$\epsilon(t)$ couples to the transition dipoles $\op{\mu}_{ij}$. 
The indexes $i,j,k$ each run over $0,1,a$.
The Hamiltonian, Eq.~\eqref{eq:full_hamil},
is represented on a Fourier grid with variable step
size \cite{SlavaJCP99,WillnerJCP04,ShimshonCPL06}.

A two-qubit gate is successfully implemented if the four basis states
$\ket{00}$, $\ket{01}$, $\ket{10}$, $\ket{11}$ transform according to
the desired unitary transformation $\op{O}$. Initially, the wavefunction for the
motional degree of freedom is given in terms of two atoms in the
ground state of the displaced harmonic oscillators,
$\varphi_0(R)=\langle R\ket{\varphi_0}$. 
Hence we consider four initial states given by
$\ket{ij(R)}=\ket{ij}\otimes\ket{\varphi_0}$, $i,j=0,1$.
The dynamics induced by a laser pulse will populate other states in
the full $3\times 3\times N_R$-dimensional Hilbert space (with $N_R$ the
number of grid points to represent the motional degree of freedom);
that is, it will induce internuclear motion leading out of the logical
subspace. In order to calculate the gate operation in the
logical subspace at the end of the pulse, 
the final states are reduced to the logical basis by tracing out the
motional degree of freedom after projection onto $\ket{\varphi_0}$.

\subsection{Optimal Control Theory for a Two-Qubit Gate}
\label{subsec:oct}

In order to implement the target operation defined in
Eq.~(\ref{eq:target_operator}), a suitable pulse $\epsilon(t)$ must be
found that drives the system evolution 
such that $\op{U}(T,0;\epsilon)= \op{O}$.
Optimal control treats 
the fidelity $F$, $F \in [0,1]$, which
measures how close the evolved two-qubit basis states $\op{U}
(T,0;\epsilon)\ket{ij(R)}$ come to the 
desired target states $\op O\ket{ij(R)}$, as a functional
of the control field $\epsilon$. Allowing for additional constraints
such as minimization of the integrated pulse energy leads to the total
functional $J$, 
\begin{equation}
  J = - F + \int_{0}^{T} g(\epsilon) \dd t \,,
  \label{eq:j_functional}
\end{equation}
which is to be minimized. Variation of the functional $J$ with respect to
the evolving two-qubit basis states and the control field yields a set of
coupled optimization equations that need to be solved iteratively
\cite{Tannor92,SomloiCP93,ZhuJCP98,PalaoPRA03}. 
We use the linear Krotov algorithm as outlined in Ref.~\cite{PalaoPRA03} to
perform the optimization. Starting with an arbitrary guess pulse
$\epsilon^{(k)}(t)$, $k=0$, the algorithm sequentially updates the pulse
at every point in time to yield an optimized pulse
$\epsilon^{(k+1)}(t)$ that is \textit{guaranteed} to improve the
target functional $J$ in each step of the iteration $k$.
We take the running cost $g(\epsilon)$ in Eq.~(\ref{eq:j_functional})
to be the change in integrated pulse energy, 
\begin{equation}
  g(\epsilon) = \frac{\alpha}{S(t)} \left[
    \epsilon^{(k+1)}(t) - \epsilon^{(k)}(t)
  \right]^2 \,,
  \label{eq:running_cost}
\end{equation}
rather than the integrated pulse energy itself.
This guarantees that, as we approach the optimum, $g(\epsilon)$ goes
to zero such that the minimum of $J$ becomes equal to the maximum of
$F$ \cite{PalaoPRA03}. In Eq.~\eqref{eq:running_cost}, $\alpha$ 
denotes an arbitrary positive scaling parameter, and the shape
function $S(t)$, 
\begin{equation}
  S(t) = \sin^2 \left( \pi t/T \right) \,,
\end{equation}
ensures that the pulse is switched on and off smoothly. 

A possible starting point to define the fidelity is given by the
complex-valued inner matrix product \cite{PalaoPRA03}, 
\begin{equation}
  \label{eq:tau}
  \tau = \sum_{i,j=0,1} \opBraket{ij(R)}{\op{O}\daggered
    \op{U}}{ij(R)} \,.
\end{equation}
Out of the several choices for obtaining a real-valued functional,
we employ 
\begin{equation}
  \label{eq:Fre}
  F = \frac{1}{N} \mathfrak{Re}\left[ \tau\right]\,,
\end{equation}
which is sensitive to a global phase \cite{PalaoPRA03}.
The Hilbert space of the optimization problem can then be reduced to
the $8 N_R$-dimensional subspace since the $\ket{11}$ level has no
coupling to any other level, see Fig.~\ref{fig:2q_levels}. 
The evolution of the $\ket{11(R)}$ state cannot be
controlled by the laser pulse but it is known to be
\begin{equation*}
  \op{U}(T,0;\epsilon) \ket{11(R)} = e^{i \phi_{T}} \ket{11(R)}; \quad
  \phi_{T} = E_{11} T/\hbar\,.
\end{equation*}
Including the information about the phase $\phi_{T}$ explicitly as a
global phase for all target states $\op O\ket{ij(R)}$, 
the evolution of \ket{11(R)} can be omitted from the system
dynamics. Since entanglement is generated only by the dynamics of the
states in the left-most column of Fig.~\ref{fig:2q_levels}, and the
remaining levels remain uncoupled from the vibrational dynamics,
the system dynamics can be further separated into those of the four
molecular states in the left-most column of Fig.~\ref{fig:2q_levels}
and those of a two-level system representing the states in the middle
columns of Fig.~\ref{fig:2q_levels}. Care must then be taken to
extract the true two-qubit phase from the evolution of the \ket{00(R)}
state which contains two-qubit and single-qubit contributions.
This is described in Appendix~\ref{app:reduction}. 
The numerical results presented below are obtained both within the
$8N_R$-dimensional model and the $4N_R$-dimensional model plus
two-level system, and the optimal pulses have been cross-checked.

For the fidelity defined in Eq.~(\ref{eq:Fre}) and the running cost
given in Eq.~(\ref{eq:running_cost}), the optimization equations read 
\cite{PalaoPRA03} 
\begin{widetext}
\begin{eqnarray}
  \Delta \epsilon(t) &=&   
  \epsilon^{(k+1)}(t) -  \epsilon^{(k)}(t) = 
  \frac{S(t)}{2\alpha} \mathfrak{Im} \left[ \;
    \sum_{i,j=0,1} 
    \opBraket{\Psi^{bw}_{ij}(R;t)}{\op{\mu}_{ij}(R)}{\Psi^{fw}_{ij}(R;t)}
    \; \right]\,,
  \label{eq:oct_pulseupdate} \\
  \bra{\Psi^{bw}_{ij}(R;t)} &=& \bra{ij(R)}
  \op{O}\daggered\op{U}\daggered(t,T;\epsilon^{(k)}) \,,
  \label{eq:oct_bw_prop} \\
  \ket{\Psi^{fw}_{ij}(R;t)} &=& \op{U}(t,0;\epsilon^{(k+1)})
  \ket{ij(R)} \,.
  \label{eq:oct_fw_prop}
\end{eqnarray}
\end{widetext}
Here, $\bra{\Psi^{bw}_{ij}(R;t)}$ denotes the backward propagated 
target state $\op{O}\ket{ij(R)}$ at time $t$. The backward propagation
is carried out using the old field,
$\epsilon^{(k)}(t)$. $\ket{\Psi^{fw}_{ij}(R;t)}$ represents the forward
propagated initial state $\ket{ij(R)}$ at time $t$. The new field,
$\epsilon^{(k+1)}(t)$, is employed in the forward propagation.
$\bra{\Psi^{bw}_{ij}(R;t)}$ and $\ket{\Psi^{fw}_{ij}(R;t)}$ are
obtained by solving the time-dependent Schr\"odinger equation
numerically with the Chebychev propagator \cite{RonnieReview94}.
The time is discretized in $n_t$ steps of width $\Delta t$, between 0 and
$T$. Since no rotating wave approximation is employed, 
$\Delta t$ has to be fairly  small (0.025$\,$fs to
0.05$\,$fs). 

For the desired gate implementation, 
there are two aspects to the optimization problem: On one hand, 
the two-qubit phase $\chi$ has to be realized. This is possible due to
the interaction between the two qubits in the electronically excited
auxiliary state. On the other hand, control over the motional degree of freedom
has to be exerted. That is, at the end of the gate operation, the
motional states $\ket{ij(R)}=\ket{ij}\otimes\ket{\varphi_0}$ have to be
fully restored, except for the phases $\phi_{ij}$. Final
wavefunctions containing contributions from eigenstates other than
$\ket{ij}\otimes\ket{\varphi_0}$ imply leakage from the quantum
register. These two aspects of the optimization result can be
quantified independently, and allow for a more thorough analysis of
the solutions to the control problem than just the fidelity.

The success of control over the motional degree of freedom for the ground state
is measured by projecting the final state, $\op U(T,0;\epsilon^{opt})
\ket{00(R)}$, onto the desired state,
\begin{equation}
  \label{eq:F00}
  F_{00} = \absSq{\opBraket{00(R)}{\op U(T,0;\epsilon^{opt})}{00(R)}}\,.
\end{equation}
The phase acquired by each of the propagated basis states is given by
\begin{equation}
  \phi_{ij} = \arg \left(
    \opBraket{ij(R)}{\op U(T,0;\epsilon^{opt})}{ij(R)} \right)\,.
\end{equation}
The phases $\phi_{ij}$ contain both single-qubit and two-qubit
contributions. This is due to the weak molecular interaction which
corresponds to small detunings of the laser from the atomic transition
line. The pulse therefore drives single-qubit purely atomic local 
transitions in addition to true two-qubit nonlocal transitions into the
molecular state. The Cartan decomposition of a two-qubit unitary
into local and non-local contributions provides a tool to extract the
desired non-local phase $\chi$ from the $\phi_{ij}$.
It states that any two-qubit gate can be decomposed as
$\op{U} = \op{k}_{1} \op{A} \op{k}_{2}$,
where $\op{k}_{1}$ and $\op{k}_{2}$ are local operations acting 
on the single qubits, and $\op{A}$ is 
a non-local operator. The non-local contribution is characterized by
local invariants~\cite{zhang2003}. 
For the phasegate considered here, the local invariants are calculated
according to Ref. \cite{zhang2003} to be 
\begin{eqnarray*}
  G_{1}\left( \op{U} \right)
  &=& \cos^{2}\left[
    \frac{1}{2}( \phi_{00} - \phi_{01} - \phi_{10} + \phi_{11}) 
  \right] \,, \label{eq:g1_diaggate}\\  
  G_{2}\left( \op{U} \right)
  &=& 2 + \cos\left[
    \phi_{00} - \phi_{01} - \phi_{10} + \phi_{11}
  \right]\,. \label{eq:g2_diaggate}
\end{eqnarray*}
On the other hand, for the desired target gate $\op{O}$ defined in
Eq.~\eqref{eq:target_operator}, the local invariants are given by
\begin{eqnarray*}
  G_{1}\left( \op{O} \right)
  = \cos^{2}\left[
    \frac{\chi}{2} 
  \right] \,, \quad
  G_{2}\left( \op{O} \right) =
  2 + \cos\left[ \chi  \right]\,. 
\end{eqnarray*}
We can thus identify the non-local phase,
\begin{equation}
  \chi = \phi_{00} - \phi_{01} - \phi_{10} + \phi_{11} \,.
  \label{eq:true_phase}
\end{equation}
The  non-local character of the implemented gate \op{U}
can also be measured in terms of the entangling 
power or concurrence~\cite{KrausPRA01}. For the controlled
phasegate it is obtained from the non-local phase $\chi$,
\begin{equation}
  C = \left| \sin \frac{\chi}{2} \right|.
  \label{eq:pg_concurrence}
\end{equation}
In the presentation of our numerical results below we will use the
motional purity $F_{00}$ and the non-local phase $\chi$
in addition to the fidelity $F$ to analyze the performance of the
optimal pulses.

\section{Results} 
\label{sec:results}

In order to obtain a clear physical picture of the limiting factors that
influence the speed of the two-qubit gate operation, we start by exploring
a regime in which the atom-atom interaction would be so strong to yield a time
scale shorter than any other in the problem. This regime is experimentally
unfeasible with optical potential both in terms of length scales and of
confinement strengths, and in this sense the calculation represents little more
than a toy model. Nevertheless, we simulate the dynamics taking into account in
detail the physical features of a real atomic species {\em as if} the geometry
considered were realizable in the laboratory via some trapping force. This
allows us to gain a thorough understanding of the relevant energy and time
scales, which we subsequently apply to a more realistic case of Rydberg-excited
atoms interacting at longer distances, compatible with realistic optical
potentials.

\subsection{Optimization for Two Calcium Atoms at Short Distance}
\label{subsec:Ca}

We consider two calcium atoms in an optical lattice at a distance of
$d = 5\,\nano\meter$ that will be excited into a low-lying excited
state. While such a distance is not feasible with 
the trapping techniques currently available in experiments,
larger distances do not provide a sufficient interaction strength in
the excited state to reach any significant fidelity in a reasonable
amount of time. Nevertheless, this unrealistic assumption allows us to
determine the physical mechanisms that limit the gate operation time. 

At close distance, the ground state wave functions of
the two atoms in a harmonic trap can have a significant overlap,
\begin{equation*}
  \braket{\Psi_{+}}{\Psi_{-}}
  \approx e ^{- \frac{m \omega d^{2}}{2 \hbar}}\,.
\end{equation*}
In order to be able to treat qubits carried by the two atoms as
independent, we compensate for the small value of $d$ with an
artificially large trap 
frequency. A choice of $\omega = 400\,\mega\hertz$ ensures that 
the overlap of the wave functions is smaller than $10^{-4}$.
The grid parameters need to be chosen such that a reasonable number
of trap eigenstates (about 50 in our case) is correctly represented.
This is accomplished by taking $R_{\min}$ to be $5.0\,$a$_0$,
$R_{\max} = 300.0\,$a$_0$, the number of grid points $N_R=512$,
with mapping parameters $\beta = 0.5$, $E_{\max} = 1 \times 10^{-8}$,
cf. Ref.~\cite{SlavaJCP99}.

The minimum gate operation time to achieve a high-fidelity
implementation can be due to (i) the strength of the exchange
interaction in the excited state, or (ii) the vibrational 
motion in the trap. We investigate both hypotheses.
The time scale associated with the interaction strength is estimated 
from the maximum phase that can accumulate in the interacting state
during time  $T$. For a non-local phase of one radian, we find
\begin{equation}
  T^{rad}_{int}(d) = \frac{1}{E_{0a} - V_{0a}(d)}\,,
\end{equation}
where $E_{0a}$ denotes the energy of two infinitely separated,
i.e. non-interacting atoms in the \ket{0a} or $\ket{a0}$
state, and $V_{0a}(d)$ the interaction potential at distance $d$. For
$d = 5\,\nano\meter$, this yields $T^{rad}_{int} \approx
1.23\,\pico\second$ for a non-local phase of one radian, and 
$T_{int}^{\pi} \approx 4.4\,\pico\second$ for a 
non-local phase of $\pi$.
%
The time scale associated with the vibrational motion in the trap is
estimated by considering the mean energy difference of the trap ground
state energy to its neighboring levels, i.e. the last bound state and
the first excited trap state. For the chosen trap frequency, we obtain
$T_{v} \approx 800\,\pico\second$.

\begin{figure}[tb]
  \centerline{
    \includegraphics{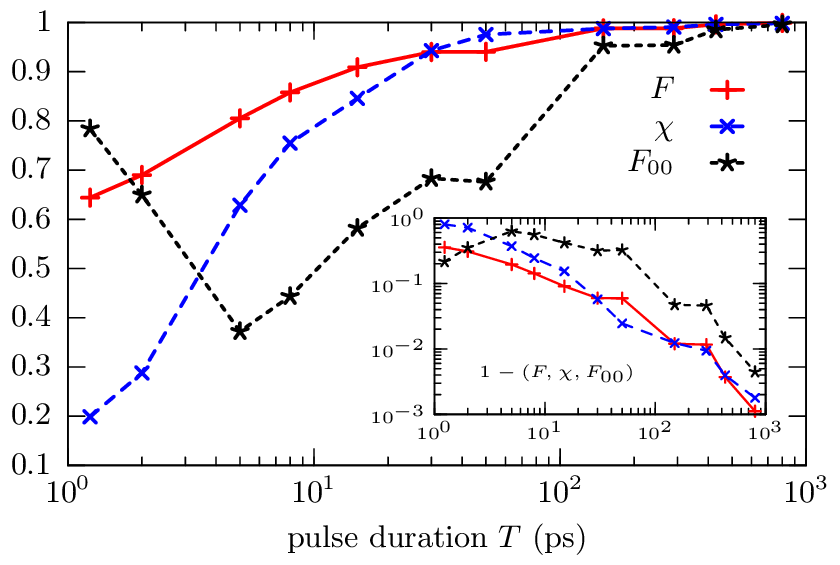}
  }
  \caption{(color online) Fidelity $F$, non-local phase $\chi$ (in units of $\pi$)
    and vibrational fidelity, i.e.,
  projection onto the vibrational target state, $F_{00}$, for different gate
  times $T$. The inset shows the infidelity $1-F$ and the respective quantities
  $1-\chi$ and $1-F_{00}$.The interatomic
  distance is $d=5\,\nano\meter$.}
  \label{fig:table1}
\end{figure}

The optimization results for gate operation times varied between the
two limits $T^{rad}_{int}$ and $T_{v}$ are shown in Figure~\ref{fig:table1}.
We compare fidelity $F$, Eq.~\eqref{eq:Fre}, non-local phase $\chi$,
Eq.~\eqref{eq:true_phase}, 
and vibrational fidelity $F_{00}$, Eq.~\eqref{eq:F00}.
The optimizations are converged to within $\Delta F < 1\times10^{-4}$
except for $T = 30~\pico\second$ and $50~\pico\second$ which are
converged to within $\Delta F < 2\times10^{-4}$.
For durations below $150\,\pico\second$, with errors remaining larger
than $10^{-2}$ no satisfactory fidelity is obtained. 
As the gate operation time approaches $T_{v}$, 
optimization is successful in the sense that fidelities
arbitrarily close to one can be reached. 
The results shown in Fig.~\ref{fig:table1} can be understood as
follows: The two-qubit phase $\chi$ increases with the
pulse duration $T$, and at $T=5\,$ps, the  time that was roughly  
estimated to reach a non-local phase of $\pi$, about half that phase is
actually obtained. This is not surprising since the wavepacket is not
in the excited state for the complete gate duration $T$
due to the switch-on and switch-off phases of the pulse and its general
shape. The non-local phase reaches the 
desired value of $\pi$ at about $50\,\pico\second$. We thus find that
a prolonged action of the exchange interaction leads indeed to a
non-local gate. However, for short gate durations, no control
over the motional degree of freedom can be exerted. For $T=5\,$ps, the
vibrational fidelity drops below 50\%, and it increases rather slowly
for larger $T$. This is due to 
the wave packet spending enough time in the excited
state to be accelerated by the $1/R^3$ potential. When the laser pulse
returns the wavepacket to the electronic ground state, it has acquired
significant vibrational energy. Since the pulse is too short to
resolve the vibrational motion in the trap, optimization cannot
identify the desired trap state and thus it cannot
counteract the excitation. Population of excited trap states after the
gate can be avoided only once the pulse is long enough to resolve
different trap states. As the gate duration becomes
comparable to  $T_{v}$, fidelities close to one are obtained. 

\begin{figure}[tb]
  \centerline{
    \includegraphics{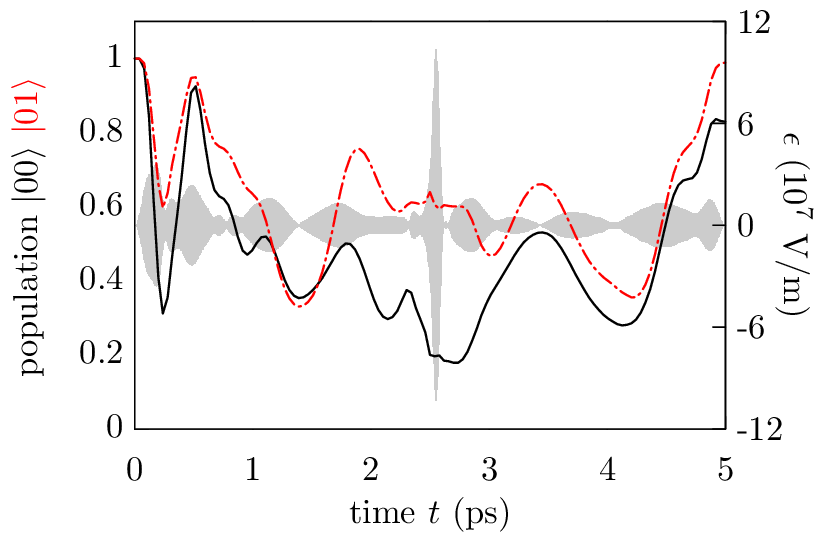}
  }
  \centerline{
    \includegraphics{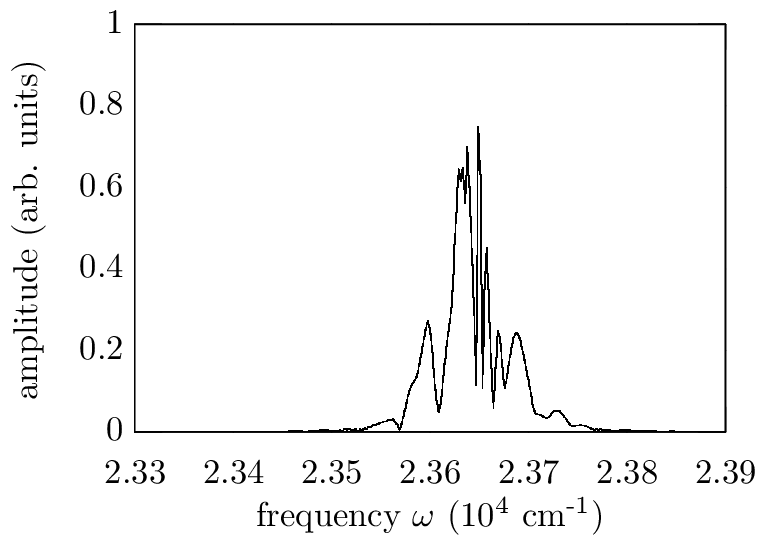}
    \hskip0.8cm 
  }
  \caption{(color online) Optimized pulse (gray) and population
    dynamics (\ket{00} state: solid black line, \ket{01} state:
    dot-dashed 
    red line, top) and pulse spectrum (bottom) for $T = 5\,\pico\second$
    ($F=0.805$).
  }
  \label{fig:5ps_pulse_pop}
\end{figure}

\begin{figure}[tb]
  \centerline{
    \includegraphics{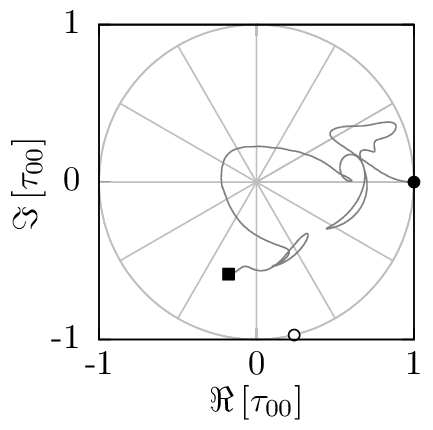}
    \includegraphics{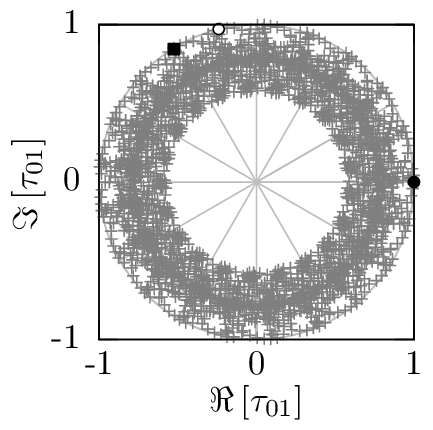}
  }
  \centerline{
    \includegraphics{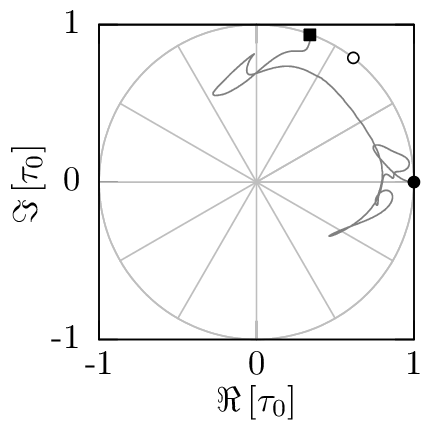}
    \includegraphics{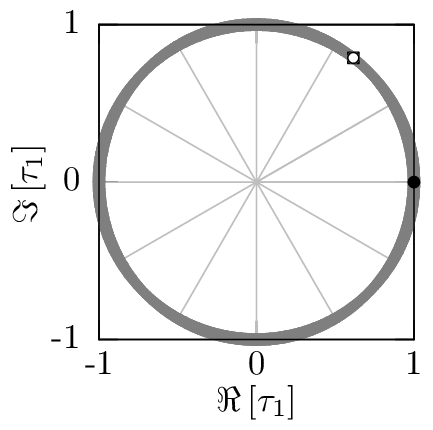}
  }
  \caption{Phase dynamics induced 
    by the optimized pulse ($T =
    5\,\pico\second$) in the complex plane for the two-qubit
    states (top) and the single-qubit states (bottom). The initial state is
    represented by a black circle, the final state by a black square, and the
    target state by a blank circle.
    }
  \label{fig:5ps_phasedyn}
\end{figure}

An analysis of the dynamics induced by the
optimized pulses  is instructive for short gate durations despite the
low fidelity.  
Figures~\ref{fig:5ps_pulse_pop} and \ref{fig:5ps_phasedyn}
display the
optimized pulse, its spectrum, and the dynamics for 
$T=5\,\pico\second$. 
The guess pulse that is used to start the iterative optimization has a
Gaussian envelope with a peak intensity of about
$4.9 \times 10^{7}\,$V/m. The intensity was chosen to drive one
complete Rabi cycle for a single qubit in the $|0\rangle$ state
($2\pi$-pulse). 
The pulse fluence is increased by a factor of about $7$ during the
course of iterations. 
Optimization results in a pulse shape that clearly shows 
more features than a Gaussian, cf. top panel of
Fig.~\ref{fig:5ps_pulse_pop}. The first peak of the pulse, centered
around $\approx 100\,$fs, drives significant population transfer to the
auxiliary state. The last peak of the pulse, centered around 
$\approx 1\,$ps, restores all population to the electronic ground state.
The dynamics yielding the non-local phase occur at intermediate
times. Since there is no interaction between the two atoms in the
$\ket{01}$ and 
$\ket{a1}$ states, the population dynamics of the $\ket{01}$ state
(red dot-dashed line in Fig.~\ref{fig:5ps_pulse_pop})
are equivalent to those of a single atom. Comparison of the 
population dynamics of the \ket{00} state and the \ket{01} state
(black solid and   red dot-dashed lines in the top panel of
Fig.~\ref{fig:5ps_pulse_pop}) therefore yields the difference
between single-qubit and two-qubit dynamics. For the short gate
operation time shown in Fig.~\ref{fig:5ps_pulse_pop}, the two
curves are fairly similar. This is in agreement with the 
non-local phase of only $0.63\,\pi$ that is achieved. 
The spectrum of the optimized pulse, cf. bottom panel of
Fig.~\ref{fig:5ps_pulse_pop}, is 
tightly centered around the $\ket{0} \rightarrow \ket{a}$ transition
frequency of 23652.30$\,$cm$^{-1}$. It is sufficiently narrow to
guarantee that no undesired transitions, for example into the
$^1\Pi_g(S+D)$ state, are induced.

The dynamics of the two-qubit system and of a single qubit can be
analyzed by projecting the time-evolved two-qubit basis state onto the 
initial two-qubit and single-qubit states,
\begin{eqnarray}
  \label{eq:taus}
  \tau_{ij}(t) &=&\opBraket{ij(R)}{\op U(t,0;\epsilon^{opt})}{ij(R)}\,,
  \\
  \tau_{j}(t) &=&\opBraket{j}{\op U(t,0;\epsilon^{opt})}{j}\,. 
\end{eqnarray}
The phase dynamics $\tau_{ij}(t)$ and $\tau_j(t)$ obtained with the
optimized pulse of Fig.~\ref{fig:5ps_pulse_pop} are shown in
Fig.~\ref{fig:5ps_phasedyn} with the top (bottom) corresponding to
the two-qubit (single-qubit) dynamics.
The phase of the initial state is indicated by a filled black circle,
the phase of the final state by a black square, and the phase of the
optimization target by a blank circle. 
The phase dynamics of the \ket{01} state (upper right panel of
Fig.~\ref{fig:5ps_phasedyn}, no connecting
lines are shown) consist of a mixture of the
natural time evolution and the dynamics induced by the pulse.
The \ket{1} state (lower right panel of
Fig.~\ref{fig:5ps_phasedyn}) displays \emph{only} the natural time 
evolution. Since it is not coupled to any other state by the pulse,
the population remains one, i.e., on the unit circle, at all times. 
The fact that the \ket{00} state does not return to the unit circle at the
final time $T$ indicates leakage out of the quantum register. 
The overall fidelity amounts to only $0.805$,
cf. Figure~\ref{fig:table1}, with the target phase for both the
\ket{00} state and \ket{01} state missed by almost equal 
amounts, cf. black squares and empty circles in the upper panel of
Fig.~\ref{fig:5ps_phasedyn}. 
This reflects that the optimization is balanced with respect
to all targets, i.e. the terms in the sum of the target functional,
Eqs.~\eqref{eq:tau}-\eqref{eq:Fre}, all enter with the same weight.
A comparison of the \ket{00} and \ket{0} phase dynamics (upper and
lower left panels of Fig.~\ref{fig:5ps_phasedyn}) 
illustrates how a true non-local phase is achieved, even though the 
optimization is only partially successful: Without interaction the
phase on the \ket{00} state  would evolve according to $\phi_{00} = 2
\phi_{0}$. The extent to which this is not the case demonstrates how
the interaction leads to the non-local phase.

\begin{figure}[tb]
  \centerline{
    \includegraphics{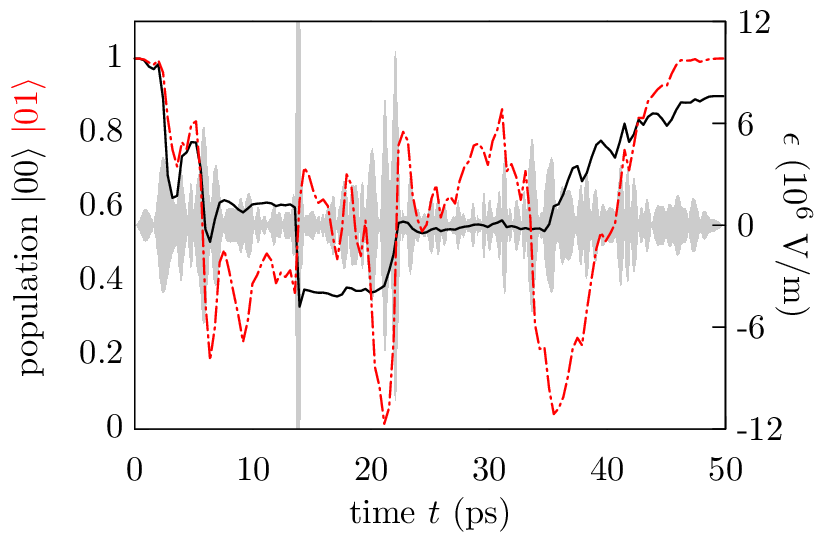}
  }
  \centerline{
    \includegraphics{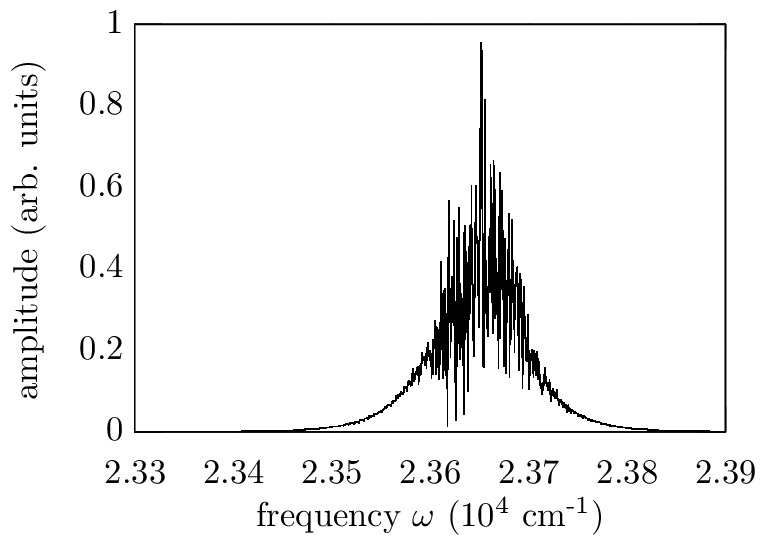}
    \hskip0.8cm 
  }
  \caption{(color online) Pulse dynamics (top) and spectrum (bottom) for the
    optimized pulse with $T = 50\,\pico\second$ after 255 iterations($F
    = 0.988$), analogously to Fig.~\ref{fig:5ps_pulse_pop}}
  \label{fig:50_ps_pulse_pop}
\end{figure}
\begin{figure}[tb]
  \centerline{
    \includegraphics{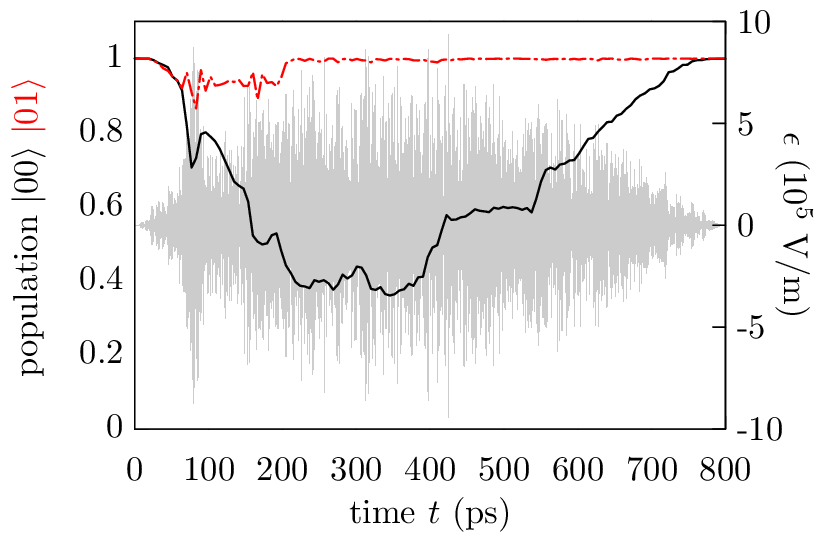}
  }
  \centerline{
    \includegraphics{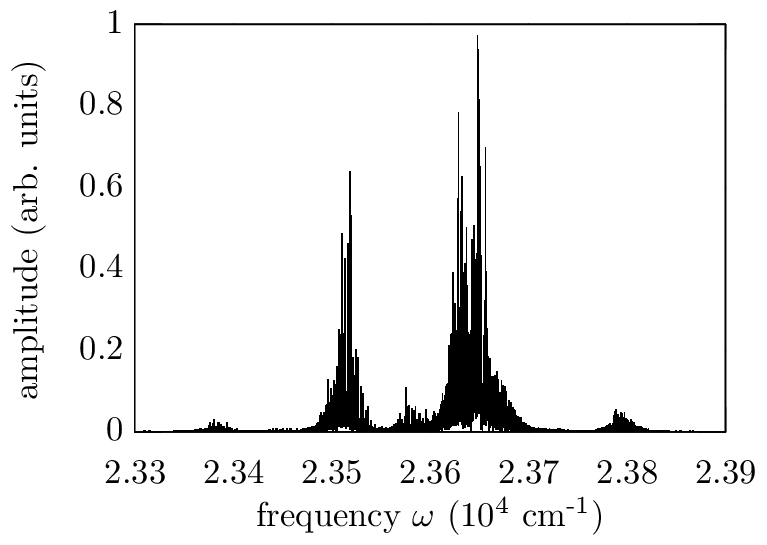}
    \hskip0.8cm 
  }
  \caption{(color online) Pulse dynamics (top) and spectrum (bottom) for the
    optimized pulse with $T = 800\,\pico\second$ after 104 iterations ($F
    = 0.999$), analogously to Fig.~\ref{fig:5ps_pulse_pop}}
  \label{fig:800ps_pulse_pop}
\end{figure}
The optimized pulses, their spectra, and the corresponding population
dynamics for intermediate and long gate durations are shown in
Figs.~\ref{fig:50_ps_pulse_pop} ($T=50\,$ps) and
\ref{fig:800ps_pulse_pop} ($T=800\,$ps). The guess pulses were again
chosen to be Gaussian $2\pi$-pulses. During the course of the
iterations, the pulse fluence was increased by a factor of 28 for
$T=50\,$ps and by a factor of 44 for $T=800\,$ps. 
The overall structure of the optimized pulse for $T=50\,\pico\second$
is similar to that obtained for $T=5\,\pico\second$: Two peaks
at the beginning and the end induce population transfer to and from
the auxiliary state while the intermediate part of the pulse drives
Rabi oscillations in the course of which the non-local phase is
achieved. The first peak of the pulse triggering population
transfer to the auxiliary state remains clearly visible
as the gate operation time $T$ is further increased,
cf. Fig.~\ref{fig:800ps_pulse_pop}. Overall, however, the optimal
pulse shows less discernible features for $T=800\,$ps than for the
shorter gate durations where a sequence of subpulses was found. This
is reflected in the population dynamics: While for $T=50\,$ps, each
subpulse drives a partial transfer, resulting in step-wise
population dynamics, an almost adiabatic behavior is observed for
$T=800\,$ps. Comparing
the population dynamics for a single qubit and the two-qubit system
for $T=50\,$ps (black solid and red dot-dashed lines in the upper
panel of Fig.~\ref{fig:50_ps_pulse_pop}), more differences are
obtained than for $T=5\,$ps, cf. Fig.~\ref{fig:5ps_pulse_pop}, but 
overall the single qubit and two-qubit dynamics are 
still fairly similar. This changes dramatically for $T=800\,$ps
(black solid and red dot-dashed lines in the upper
panel of Fig.~\ref{fig:800ps_pulse_pop}), where 
the populations dynamics for \ket{00} and \ket{01}
are clearly distinct, reflecting that  the desired non-local phase 
is fully achieved ($\chi=0.998\pi$ for $T=800\,$ps as compared to 
$\chi=0.975\pi$ for $T=50\,$ps).
The spectrum of the optimal pulse for $T=50\,$ps is fairly similar to
that obtained for $T=5\,$ps, cf. the lower panels of
Figs.~\ref{fig:5ps_pulse_pop} and \ref{fig:50_ps_pulse_pop}: It
basically consists of a single narrow peak centered around the
$\ket{0}\to\ket{a}$ transition frequency. The spectrum for $T=50\,$ps
shows somewhat more features within the peak which is attributed to
the better spectral resolution for larger $T$.
For $T=800\,$ps, the spectrum of the
optimal pulse consists of a narrow peak at the $\ket{0}\to\ket{a}$
transition frequency and sidebands. These sidebands remain
sufficiently close to the $\ket{0}\to\ket{a}$ transition, that
resonant excitation into other electronic states can be excluded,
cf. Fig.~\ref{fig:ca2pot}. 
\begin{figure}[tb]
  \centerline{
    \includegraphics{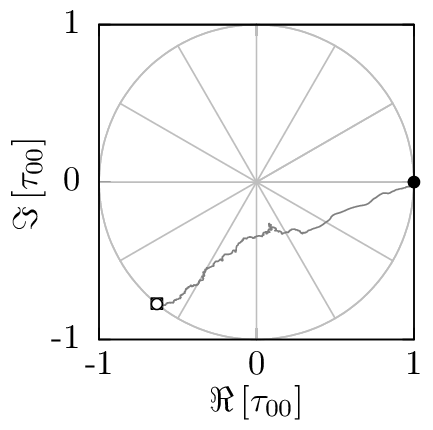}
    \includegraphics{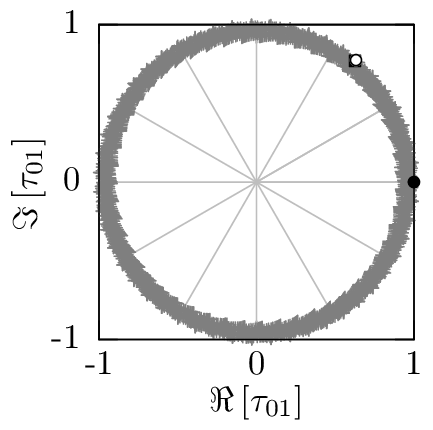}
  }
  \centerline{
    \includegraphics{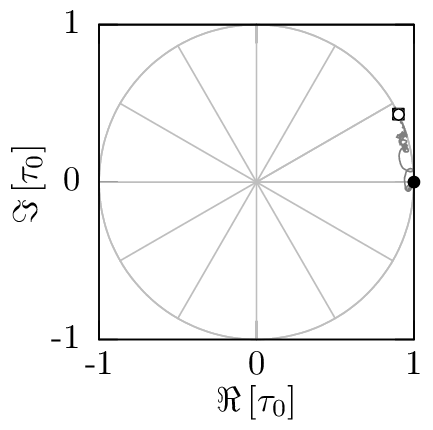}
    \includegraphics{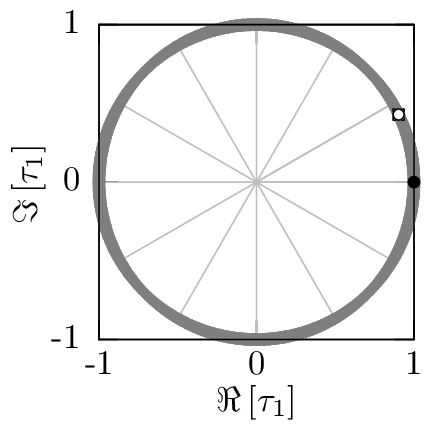}
  }
  \caption{Phase dynamics induced by the optimized pulse ($T =
    800\,\pico\second$) in the complex plane for the two-qubit
    single-qubit states,
    analogously to Fig.~\ref{fig:5ps_phasedyn}.}
  \label{fig:800ps_phasedyn}
\end{figure}
The phase dynamics induced by the optimized pulse of
Fig.~\ref{fig:800ps_pulse_pop} are shown in Fig.~\ref{fig:800ps_phasedyn}.
Since the target phases include the natural time evolution, their
locations in Fig.~\ref{fig:800ps_phasedyn} differ from those in
Fig.~\ref{fig:5ps_phasedyn}. The overlap of the final states (black
square) and the target states (open circle) confirms success of the
optimization. All phases end up on the unit circle demonstrating that
no leakage from the quantum register occurs at the end of the gate for
$T=800\,$ps. 

We also carried out optimizations for non-local target phases that are
a fraction of $\pi$ such as $\frac{\pi}{2}$ or $\frac{\pi}{3}$. If
high-fidelity implementations of such fractional phasegates are found, 
several of these gates can be combined sequentially to yield
a total non-local phase of $\pi$. 
However, for short gate operation times, optimization for
non-local target phases smaller than $\pi$ did not prove any more
successful than optimization for $\pi$. 
In particular, population of excited trap states at the end of the
gate could not be avoided for fractional phasegates either.
Moreover, we investigated
whether pulses driving multi-photon transitions, for example pulses
with their central frequency a third of $\ket{0} \rightarrow
\ket{1}$ transition frequency, 
yield better fidelities for short gate operation times. However, we
did not observe any substantial difference in the results compared to
the pulses reported in Figs.~\ref{fig:5ps_pulse_pop},
\ref{fig:50_ps_pulse_pop},  and \ref{fig:800ps_pulse_pop}.
These additional investigations confirm that for our example of two
ultracold calcium atoms in an optical lattice, the limits on the gate
operation time is set by the requirement to restore the ground
vibrational state of the trap. 

\subsection{Optimization for Two Atoms at Long Distance under 
  Strong Dipole-Dipole Interaction}
\label{subsec:dipole}

To determine whether it is really the ground state motion in the trap
and not the non-local interaction in the excited state
that sets the speed limit for two atoms resonantly excited to an
interacting state, we vary the interaction strength $C_3$ of the 
dipole-dipole interaction potential,
\begin{equation}
  \op{V}(R)_{0a} = \op{V}(R)_{a0} = - \frac{C_3}{R^{3}}\,,
\end{equation}
keeping the trap frequency constant.
We consider the atoms to be separated by $d = 200\,\nano\meter$ which 
corresponds to a realistic optical lattice in the UV regime. 
In order to keep the overlap of the ground state wave functions smaller than
$10^{-4}$ at a distance of $200\,\nano\meter$ the trap frequency has
to be set to at least $250\,\kilo\hertz$. This corresponds to
$T_{v}\approx 2\,$ns. 

For the interaction potential of two calcium
atoms in the $B^1\Sigma_u^+$ state used in 
Sec.~\ref{subsec:Ca}, the $C_3$ coefficient takes a value of 
$16.04\,$a.u.$=0.5217\times
10^3\,$nm$^3$cm$^{-1}$~\cite{BusseryPRA03,BusseryMolPhys06,DegenhardtPRA03}. 
This results in an interaction energy of about $4\,\wavenumbers$ at $d 
= 5\,\nano\meter$. Based on the results of Sec.~\ref{subsec:Ca}, we
know that such an interaction energy is sufficient to yield a
non-local phase in a few tens of picoseconds. For $d=200\,$nm, 
the same interaction energy is obtained by choosing 
$C_{3}$ to be roughly $1\times 10^{6}\,$a.u. Just for comparison, the
$C_3$ coefficient for highly excited Rydberg states is about
$3\times 10^{6}\,$a.u., resulting in an interaction energy of about
$1.3\times 10^{-3}\,\wavenumbers$  at a typical distance of
$4\,\mu\meter$ for two atoms trapped in optical tweezers
\cite{GaetanNaturePhys2009}. 

We vary the $C_3$ coefficient from $1 \times 10^{6}\,$a.u. to $1
\times 10^{9}\,$a.u.  If the gate duration is solely determined by the
requirement of a sufficiently strong interaction to
realize the non-local phase, we expect to find high-fidelity
implementations with optimal control by increasing the $C_3$
coefficient. In particular, we pose the question whether 
picosecond and sub-picosecond gate durations can be achieved given
that the interaction is sufficiently strong, i.e. given that the $C_3$ 
coefficient is sufficiently large. Based on the results of
Sec.~\ref{subsec:Ca} where a non-local phase of $\pi$ was achieved
within 50$\,$ps, we estimate that $C_3$ needs to be increased 
from $1 \times 10^{6}$ by a factor of 50 (100) to obtain a
high-fidelity gate for a duration of $1\,$ps ($0.5\,$ps).

\begin{figure}[tb]
  \centerline{
    \includegraphics{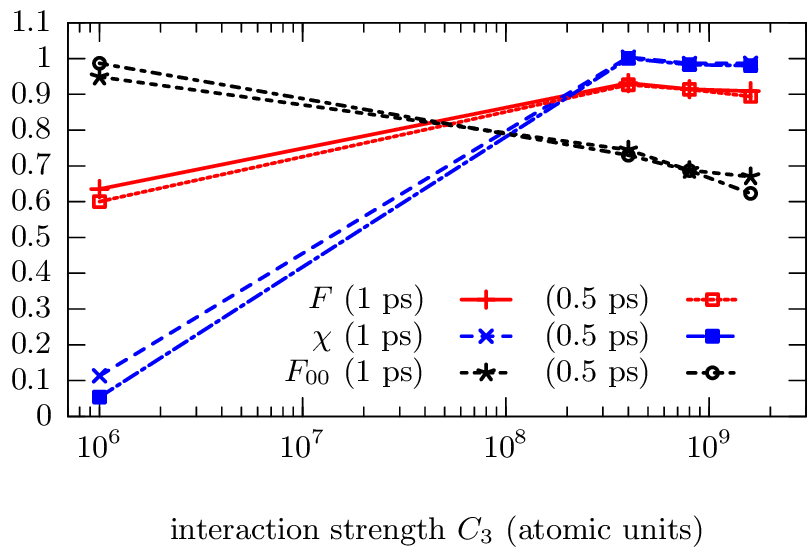}
  }
  \caption{(color online)
    Fidelity $F$, non-local phase $\chi$ and vibrational fidelity, i.e.
    projection onto the vibrational target state, $F_{00}$ for increasing
    interaction strength $C_3$ in the excited state for two different gate
    times $T$. All optimizations have converged to $\Delta F < 1 \times
    10^{-4}$ The interatomic distance is $d=200\,\nano\meter$.
  }
  \label{fig:table2}
\end{figure}

Figure~\ref{fig:table2} presents optimization results for 
a controlled phasegate with gate operation times of $T =
0.5\,\pico\second$ and $T = 1\,\pico\second$.
The central frequency of the guess
pulse was adjusted in each case to compensate for the increased interaction
energy and ensure resonant excitation. 
The grid parameters were chosen to be $R_{\min} = 5\,$a$_0$, $R_{\max}
= 13000\,$a$_0\approx 688\,$nm, and $N_{R} = 2048$. This choice of $R_{\max}$
guarantees that at least fifty eigenstates of the trap are accurately
represented.  We verified that the grid is sufficiently large, i.e.,
the wave packet does not reach the boundaries of the
grid during propagation. 
Moreover we checked that doubling the number of grid points did
not yield substantially different results.
Figure~\ref{fig:table2} clearly shows that increasing the
interaction strength leads to larger non-local phases. A non-local
phase of $\pi$ is reached for $C_3=4\times 10^{8}\,$a.u.
However, increasing the interaction strength also results in a complete
loss of control 
over the motional degree of freedom, with the vibrational fidelity
$F_{00}$ reduced to 75\% and below. A stronger interaction in the
excited state accelerates the wave packet more, increasing its
vibrational excitation. This results in a larger spread over many trap
states upon the wave packet's return to the ground state. Since
$0.5\,$ps or $1\,$ps are much, much shorter than the time scale of the
motion in the trap, $T_v$, the optimization algorithm cannot resolve
the eigenstates of the trap. Thus it cannot identify the target
$F_{00}$ which is consequently missed completely. Note that trap
frequencies up to a few MHz are possible and imply shorter
$T_v$. Nevertheless, any realistic trap frequency results in
vibrational time scales much larger than a few picoseconds. While
excited state potentials providing for a strong interaction between
two neutral atoms exist, resonant excitation into such an excited
state will not yield an ultrafast non-local gate due to coupling with
the motional degree of freedom, unless we consider a gate scheme that is
completely insensitive to vibrational excitations

\section{Summary and conclusions} 
\label{sec:concl}

We have studied high-fidelity implementations of a controlled
phasegate for two trapped ultracold atoms via resonant optical
transitions to an electronically excited state with long-range 
diatomic interaction. 
To the best of our knowledge, we have for the first time 
explicitly accounted for the detailed $R$-dependence of the interaction and
thus for the coupling between electronic and nuclear dynamics 
that may cause leakage out of the quantum register.
We have employed optimal control theory to calculate laser pulses
that carry out the gate. This has allowed us to determine gate
implementations of basically arbitrarily high fidelity provided
the gate operation time is sufficiently long (and at the same time
short enough to neglect dissipation). Our main goal was to achieve the
fastest possible gate implementation and to identify what limits the
gate operation time, i.e., to determine the quantum speed limit for a 
controlled phasegate for two neutral trapped atoms. 

The standard reasoning considers the interaction strength to be the
limiting factor, i.e., the gate operation time is estimated by the
inverse of the two-qubit interaction. 
Our calculations show that a second time scale might come into play:
For resonant excitation, the interaction
between the two atoms causes a coupling between electronic and
nuclear dynamics. This induces vibrational excitation which can be
carried away by the laser pulse only if the target state is fully
resolved during the optimization. The gate operation time is thus
limited either by the two-qubit interaction strength or by the
vibrational motion in the trap, which ever one of the two yields the
larger time. 

This finding has important implications for the design of two-qubit
gates where the qubits are carried by neutral atoms. For
example, excitation of atoms into Rydberg states yields an
interaction that one might expect to allow for nanosecond to
sub-nanosecond gate operation times. However, the motional state of
the atoms needs to be restored at the end of the gate, and  
traps with sub-nanosecond vibrational motion seem difficult to realize. 
Thus the question of how an ultrafast two-qubit gate can be realized in a
scalable setup still remains  open. 

\begin{acknowledgments}
  We would like to thank Peter Zoller for many stimulating discussions.
  Financial support  from the Deutsche Forschungsgemeinschaft (Grant No. KO
  2302), as well as the BMBF (project QUOREP) and the EC (IP AQUTE) is
  gratefully acknowledged .
\end{acknowledgments}

\appendix

\section{Reduced Optimization Scheme}
\label{app:reduction}

Since the dynamics relevant for obtaining a non-local phase involves
only the $\ket{00}$ state out of the four two-qubit states,
cf. Fig.~\ref{fig:2q_levels}, we can reduce our full model,
Eq.~\eqref{eq:full_hamil}, to one describing only the left-most column
of Fig.~\ref{fig:2q_levels}. This is a direct consequence of
optimizing for a diagonal two-qubit gate. 
We are then operating in a $4N_R$-dimensional Hilbert space  
instead of a $3\times 3\times N_R$-dimensional Hilbert space.
However, care must be taken to extract the correct non-local
phase $\chi$. The phase $\phi_{00}$ describing the time evolution of
the two-qubit $\ket{00}$ state alone is not sufficient to obtain the
non-local phase $\chi$ since 
$\phi_{00}$ contains contributions from both the single-qubit and two-qubit
dynamics. We therefore need to augment our reduced model for the
dynamics starting from the $\ket{00(R)}$ by a two-level system
(\ket{0}, \ket{a}). This captures the purely single-qubit dynamics in
the phase $\phi_0$. The non-local phase is then obtained as the
difference between $\phi_{00}$ and (twice) $\phi_0$, 
see Table~\ref{tab:full_reduced_system}. The optimization targets 
for the full model and the reduced model are
also listed in Table~\ref{tab:full_reduced_system}. They correspond to
the unitary transformation for the full model and to two
state-to-state transitions for the reduced model. Note that this type
of state-to-state transition requires a phase-sensitive functional
such as the one in Eq.~\eqref{eq:Fre}.
We have checked numerically that the full and reduced model are indeed
equivalent: Propagating the Schr\"odinger equation with an optimal
pulse obtained for the reduced model but employing the full
Hamiltonian, Eq.~\eqref{eq:full_hamil}, we obtained the same fidelity
as for the reduced model.

\begin{table}[bt]
  \vskip8mm
  \begin{tabular}{|l|cc|}
    \hline 
    \parbox{1.6cm}{\vskip2ex~\vskip1ex} & full & reduced \\
    \hline 
    target           & \parbox{1cm}{
      \begin{align*}
        \ket{00} &\rightarrow
        e^{i (\phi + \phi_{T})} \ket{00} \\
        \ket{01} &\rightarrow     e^{i \phi_{T}} \ket{01} \\
        \ket{10} &\rightarrow     e^{i \phi_{T}} \ket{10} \\
        \ket{11} &\rightarrow     e^{i \phi_{T}} \ket{11}
      \end{align*}}  & \parbox{1cm}{
      \begin{align*}
        \ket{00} &\rightarrow     e^{i (\phi + \phi_{T})} \ket{00}  \\
        \ket{0}  &\rightarrow     e^{i \phi_{T} /2} \ket{0}
      \end{align*}}  \\ \hline
    gate phases      & \parbox{1cm}{
      \begin{align*}
        &\phi_{00} \\
        &\phi_{10} = \phi_{01} \\
        &\phi_{11}
      \end{align*}}  & \parbox{1cm}{
      \begin{align*}
        &= \phi_{00} \\
        &= \phi_{0} +\phi_{1} \\
        &= 2 \phi_{1}
      \end{align*}}            \\ \hline
    \parbox{1.6cm}{
    \vskip1ex
    \raggedright
    non-local \\phase
    \vskip1ex
    }
                     & $\chi = \phi_{00} - \phi_{01} - \phi_{10} +
    \phi_{11}$  & $\chi = \phi_{00} - 2 \phi_{0}$\\
    \hline
  \end{tabular}
  \caption{Comparison between the full model and the reduced model for
    optimization of a controlled phasegate.}
  \label{tab:full_reduced_system}
\end{table}

\bibliography{phasegate}

\end{document}